# Habitability and sub glacial liquid water on planets of M-dwarf stars

Amri Wandel [1]
1. Racah Institute of Physics, The Hebrew University of Jerusalem, Jerusalem, Israel

A long-standing issue in astrobiology is whether planets orbiting the most abundant type of stars, M-dwarfs, can support liquid water and eventually life. A new study shows that subglacial melting may provide an answer, significantly extending the habitability region, in particular around M-dwarf stars, which are also the most promising for biosignature detection with the present and near-future technology.

Conditions for sustained presence of liquid water, considered to be one of the basic requirements for life, may be found on exoplanets within what is known as the Habitable Zone (HZ) of their respective star, given the planet has a suitable atmosphere. For example,
the HZ of our sun extends roughly between the orbits of Venus and Mars. However, both these planets do not have liquid water on their surface because of their atmospheres: while Mars's atmosphere is too tenuous, that of Venus is too dense and consists mainly of carbon dioxide, causing a huge greenhouse effect and a very high surface temperature. Fig. 1 illustrates the boundaries of the HZ for various models and different stellar types, including sunlike (type G) stars. Nonetheless, in the solar system, liquid water also exists out of the HZ, e.g. in Enceladus and Europa, moons of Saturn and Jupiter, respectively, as a subglacial ocean due to tidal heating. Ojha et al. [1] modeled the evolution of ice sheets and the conditions that allow liquid water to be produced and maintained at temperatures above freezing on Earth-like exoplanets. Applying their results to enhance the definition of habitability by including subglacial liquid water, we suggest more extended HZ boundaries so there may exist a wider circumstellar region with a potential for subglacial liquid water and life in colder planets and their satellites outside the traditional HZ. Furthermore, inward planets that are closer to their host star than the inner boundary of the traditional HZ, which were until now considered too hot, may also, under certain conditions, support subglacial liquid water and life. Planets within the HZ of M-dwarf stars (cooler and much less luminous than our sun) or closer to their host star are likely to be tidally locked [2] and may have temperatures adequate for liquid water, at least on part of their surface, under a broad range of atmospheric conditions [3]. As M-dwarfs are the most abundant stellar type, this has a major effect on our calculations of the abundance of potentially habitable planets, i.e. small rocky planets with a potential to support liquid water and life on the surface or under an ice cover. Planets of M-dwarfs are also the most readily observable for biosignatures [2], hence those with subglacial oceans harboring life may also be most easily detected by biomarker lookup, provided the subglacial water can reach the surface, e.g. though cracks or faults in the ice cover, producing geysers or plumes (like those observed in Enceladus and Europa).

## Subglacial liquid water

Maintaining liquid water on the planet's surface for long periods, as might be required for the evolution of life, can be impossible beyond the HZ, and even within it if there's no atmosphere with a sufficient greenhouse effect such as in the case of Mars. Producing liquid water by subglacial melting of local or global ice sheets by geothermal heat may be an alternative to radiative heat from the host star. Ojha et al.[1] have shown that liquid water may be produced and maintained at temperatures above freezing even on planets with modest (as low as 0.1 Earth's) geothermal heat produced by radiogenic elements. They find that subglacial oceans or lakes of liquid water can form by basal melting and persist under ice sheets on Earth-sized exoplanets even for surface temperatures as low as 200K expected e.g. on Trappist-1g. Under the weight of high-pressure ices, the basal meltwater is buoyant and can migrate upward, feeding intra-glacial lakes and eventually a main water layer or ocean.  Basal melting is more likely to occur on planets with thicker ice sheets, higher surface gravity and higher surface temperatures.

Protected by ice layers from energetic radiation of the host star (especially ultraviolet and X-ray radiation likely to occur in outbursts during the early evolutionary stages of M-dwarfs) such subglacial oceans may give a safe neighborhood during the extended period required for the evolution of organic life. Therefore, they are especially suitable for maintaining habitable environments on Earth- to Super Earth-sized exoplanets outside the conservative Habitable Zone of M-stars. Being in contact with the crust such subglacial lakes may provide attractive prospects for the evolution of life, somewhat similar to Europa and Enceladus. For intra-glacial lakes such a chemical contact with the crust has been questioned, but recent works [4] demonstrated that salty high-pressure ices do provide sufficient vertical permeability.

In addition to heat from radioactive elements, subglacial liquid water may be produced by tidal heating. This mechanism can maintain basal melting on planets which do not have enough geothermal radioactive heat production, e.g. small or old planets, if they are closely orbiting their hosts stars. It may be particularly relevant for small (sub-Earth-sized) planets orbiting M-dwarf stars.

**Inwards and outwards of the Habitable Zone**

The HZ extends between the inner boundary (water evaporates) and the outer one (water freezes). Inward of the HZ (closer to the host star), planets of M-dwarfs (and also of cool K-type stars) are tidally locked, one side constantly facing the host star. If they are rocky and have little or no atmosphere (like Mercury) there can be a large temperature difference between the day and night sides. While the day side may be too hot for maintaining liquid water, the water on the night side is frozen, unless too much heat is transported from the day side (e.g. by atmospheric convection) evaporating the ice. If the heat transport is low, subglacial liquid water is likely to exist and provide a potentially habitable neighborhood.

Another mechanism suggested for producing ice sheets on locked planets is when water evaporates at the sub stellar side, being transported and trapped as ice on the night side [5].This mechanism can work if the geothermal heat is low enough [6]. Even in this case basal melting may be driven by tidal heating. Another obstacle to the habitability of planets of M dwarfs may be atmospheric erosion by stellar winds [7]. Being isolated from apace by the ice cover, subglacial oceans can survive atmospheric erosion and exist even without or little atmosphere.

Taking into account subglacial liquid water, the HZ may be extended also inwards. The inner extended HZ-boundary occurs when the heat transported from the day-side of the locked planet melts and eventually evaporates the ice sheets, as pointed out above. Assuming the relevant flux is set by the Moist Greenhouse limit [8], Fig. 1 shows two inner boundaries, for 30% or 50% of the radiative energy at the day side being transported to the night side. The lower the transported fraction, the closer is the inner extended HZ boundary to the host star.  Also shown in Fig. 1 are the inner HZ boundaries according to the Recent Venus hypothesis (a phenomenological value derived from the flux received relatively recently by Venus, assuming that until about a billion years ago Venus still had liquid water on its surface) and the 3D model for locked, slowly and rapidly rotating planets [9].

Outwards of the HZ subglacial oceans may exist on geothermally heated rocky exoplanets even for low surface temperatures. Subglacial liquid water may exist out of the traditional HZ also for K- and G-stars, but its contribution to the abundance of habitable planets is most significant for M-dwarfs, as they are the majority of stars, yet their planets might be endangered by the host's energetic radiation. As a conservative outer limit for the HZ extended by basal melting, one may use the flux on Trappist 1g, the coldest exoplanet calculated by Ojha et al. [1], which is 25% of the flux received by Earth (0.25 $S_0$). This value is roughly coincident with the outer boundary set by the 1D model, as well as the Early Mars hypothesis [9] (a phenomenological border derived from the flux received some 4 billion years ago by Mars, building on the evidence that early Mars did have liquid water on its surface). However, a yet lower flux value with a further outer boundary of the HZ may be derived from the evidence for an intra-glacial lake in Mars's south pole [11]. Taking for the radiative flux received by Martian polar region $\cos(i)=0.2$ gives a flux of approximately 0.1 $S_0$, which is denoted in Fig. 1 by the line marked Martian polar lakes.

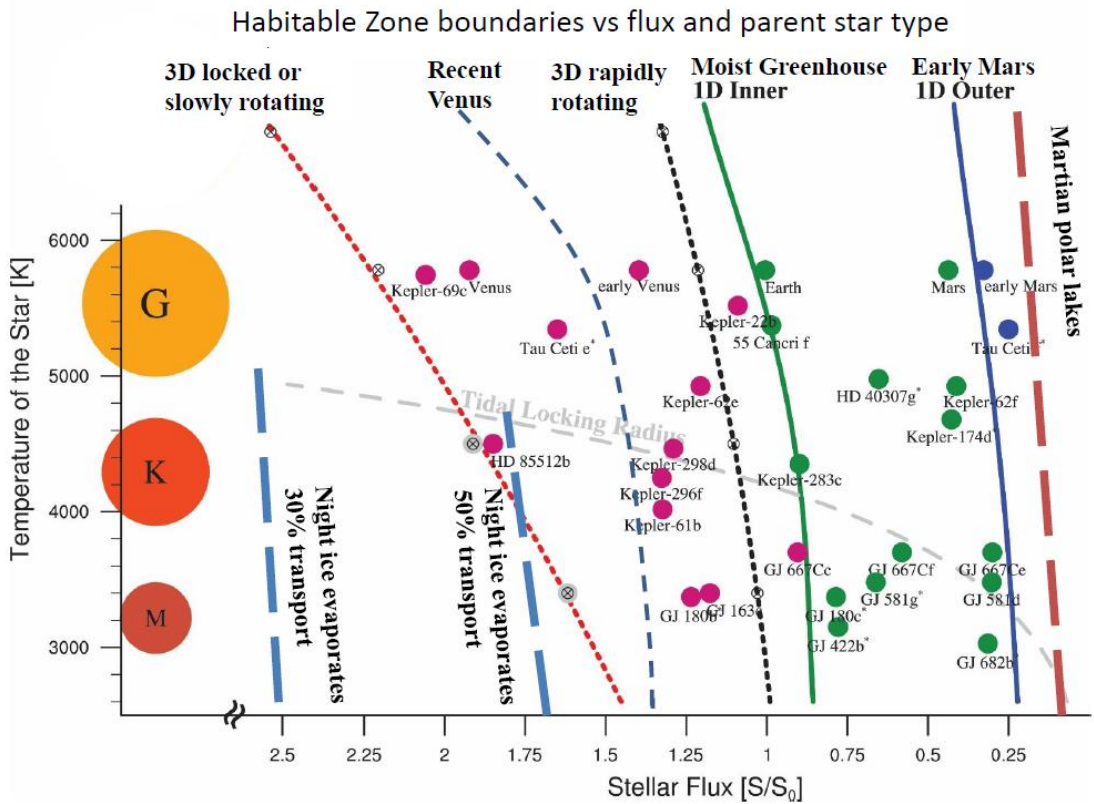

Figure 1: Habitable Zone boundaries as a function of stellar type and radiative flux received from the host star, relative to Earth. Long-dashed lines represent the extended boundaries due to subglacial liquid water suggested in this work: blue for the inner HZ boundary (night-side ice evaporates) and red for the outer one (Martian polar lakes). Green and blue lines denote the inner

(runaway greenhouse) and outer (maximum $CO_2$ greenhouse) HZ boundaries, respectively, using the 1-D model [10]. Also marked is the Recent Venus model (medium-dashed blue line), and the 3D Global Climate Model for slowly and rapidly rotating planets [9] (short-dashed red and black curves, respectively). The grey line marks the tidal locking radius [8]. The circles denote Venus, Earth and Mars as well as a few exoplanets. Adaptation from Fig. 3 of Yang et al. [9].

**The abundance of habitable planets and life-supporting ocean worlds**

Didier Queloz, the co-discoverer of the first exoplanet and Nobel laureate said [12]
"I can't believe we are the only living entity in the whole universe. There's just way too many planets, way too many stars ... the chemistry that led to life has to happen elsewhere". On that note, the extended habitability region described above, in particular around the abundant M-dwarfs, may significantly enhance the number of planets with the chemistry needed for the evolution of life. At the same time the ice sheets provide protection against the harsh energetic radiation from these stars, which is sometimes considered prohibitive for the evolution of life [3,7]. As demonstrated below, considering tidally locked planets of M-dwarfs and an extended HZ due to subglacial liquid water may increase the abundance of potentially habitable worlds by a factor of up to approximately 100.

In the solar neighborhood there is about one star per 10 light years cubed, of which 72% are M-dwarfs [13], while sunlike stars are just 4% of the stars around us and the somewhat cooler and less luminous stars of type K consist around 16%. Thus, assuming also K- and M-dwarfs can support the evolution of life increases the abundance of potentially habitable planets more than 20-fold, compared with the conservative assumption that only sunlike stars can. The Kepler mission has shown that between 10%–50% of all stars host a small, rocky planet within their habitable zone [14,15]. This considerably wide range is a result of the uncertainty on the climate model, which determines the boundaries of the Habitable Zone (Fig. 1).
Figure 2 shows the number of expected habitable planets as a function of the distance from Earth for various star types: only G-type host stars (lowest curve), G+K type stars (second lowest curve), and all three types (third curve). While the three lower curves use a conservative climate model, yielding a relatively narrow HZ (approximately the 1D model in Fig. 1), so that approximately 20% of the stars host a small rocky planet within their HZ, the upper curve assumes an extended HZ including subglacial liquid water (the thick long-dashed curves in Fig. 1), increasing the estimated percentage of stars with HZ-planets to about 100%, namely, one habitable planet per M-dwarf. Excluding binary stars, as it has been argued that the orbits of their planets may be unstable, would lower these numbers by a factor of approximately 2.

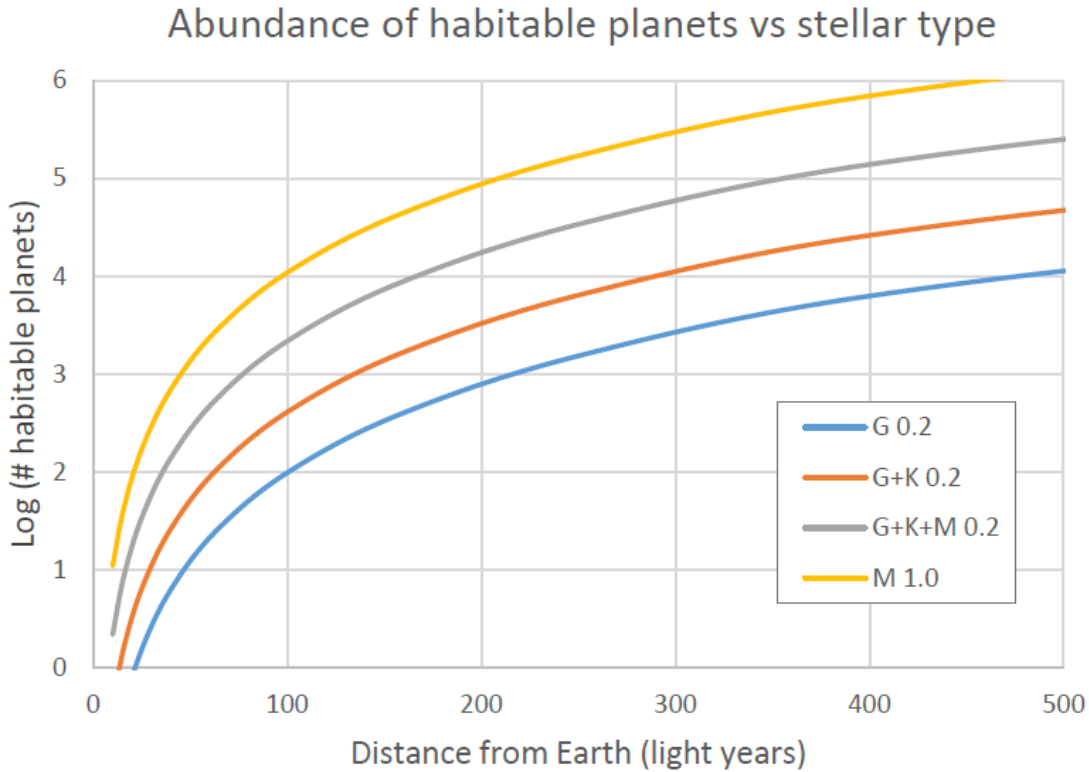

Fig. 2 The number of expected habitable planets as a function of the distance from Earth for various star types. Lower curve: only sunlike (G-type) host stars, second lowest curve: G+K type stars, third curve: G,K, and M type, all assuming 0.2 habitable planets per star. The upper curve is for M-dwarfs assuming one habitable planet per star.

To conclude, basal melting and subglacial oceans [1] combined with the extended HZ of M-dwarfs [2,16] may provide habitable environments on as much as one planet per star, compared with merely one habitable planet per 100 stars, if only G-type host stars and a conservative HZ model are considered. For closely orbiting locked planets of M-dwarfs, surface liquid water may be found on the night side which is also protected from the energetic radiation of the host star in its early evolutionary stages. On further out planets, which may be less vulnerable to radiation but too cold for surface liquid water, basal melting may produce long lasting subglacial lakes or oceans which can provide protected environment for the evolution of life outside the traditional HZ [17]. Estimating the abundance of habitable worlds, especially around M-dwarf stars is important to identifying eventual targets for biosignature research by JWST and future telescopes. In addition, it affects the likelihood and abundance of potential active extraterrestrial civilizations that should be expected according to the Drake equation and the Fermi Paradox [18].

References

1. Ojha, L., Troncone, B., Buffo, J., et al., Liquid water on cold exo-Earths via basal melting of ice sheets *Nature Comm.* **13**, 7521 (2022).
2. Shields, A. L., Ballard, S. & Johnson, J. A. The habitability of planets orbiting M-dwarf stars. *Phys. Rep.* 663, 1–38 (2016).

3. Wandel, A. On the bio-habitability of M-dwarf planets. *Astrophys. J.* **858,** 1 (2018).
4. Journaux, B., Salty ice and the dilemma of ocean exoplanet habitability *Nature Comm.* **13**, 3304 (2022).
5. Lecont, J., Forget, F., Charnay, B. et al., 3-D climate modeling of close-in land planets: Circulation patterns, climate moist bistability, and habitability. *Astron. Astrophys.* **554**, A69: 1- 17 (2013).
6. Yang, J., Liu, Y., Hu, Y. & Abbot, D. S. Water trapping on tidally locked terrestrial planets requires special conditions. *Astrophys. J. Lett.* **796**, L22 (2014).
7. J. M. Rodríguez-Mozos[1] and A. Moya Erosion of an exoplanetary atmosphere caused by stellar winds. *Astron. Astrophys.* **630**, A52 (2019).
8. Kasting, J.F., Whitmire, D.P. & Reynolds, R.T. Habitable Zones around Main Sequence Stars, *Icarus* **101**: 108-128. (1993).
9. Yang, J., Boué, G., Fabrycky, D. C., & Abbot, D. S. Strong Dependence of the Inner Edge of the Habitable Zone on Planetary Rotation Rate. *Astrophys. J. Lett.* **787**, L2 (2014).
10. Kopparapu, R. K., Ramirez, R., Kasting, J. F., et al. *Astrophys. J.* **770**, 82 (2013).
11. A Orosei, R.; et al. Radar evidence of subglacial liquid water on Mars. *Science* **361** (6401): 490–493 (2018).
12. https://unescoalfozanprize.org/professor-didier-queloz/ retrieved 24.2.2023.
13. Golovin, A.,Reffert, S., Just, A., et. al. The Fifth Catalogue of Nearby Stars (CNS5). *Astron. Astrophys.* **670A**, 19G (2023).
14. Dressing, C. D. & Charbonneau, D. The occurrence of potentially habitable planets orbiting M dwarfs estimated from the full Kepler dataset and an empirical measurement of the detection sensitivity. *Astrophys. J.* **807**, 45-67 (2015).
15. Cassan, A., Kubas, D., Beaulieu J.-P., et al. One or more bound planets per Milky Way star from microlensing observations. *Nature* **481**, 167-169 (2012).
16. Wandel, A. & Gale, J. The Bio-habitable Zone and atmospheric properties for Planets of Red Dwarfs. *Int. J. Astrobiology* **19**, 126W (2020).
17. Barge, L. M. & Rodriguez, L. E. Life on Enceladus? It depends on its origin. *Nat. Astron.* **5**, 740–741 (2021).
18. Wandel, A. The Fermi Paradox Revisited: Technosignatures and the Contact Era *Astrophys. J.* **941,** 184 (2022).

## Authors Contributions

A.W. conceived the extension of the HZ using subglacial liquid water and wrote the manuscript.

## Corresponding author

Correspondence to Amri Wandel (amri@huji.ac.il)

## Competing interests

The author declares no competing interests.